# Quantum Cryptography in Practice

Chip Elliott
BBN Technologies
10 Moulton Street
Cambridge, MA  02138

celliott@bbn.com

Dr. David Pearson
BBN Technologies
10 Moulton Street
Cambridge, MA  02138

dpearson@bbn.com

Dr. Gregory Troxel
BBN Technologies
10 Moulton Street
Cambridge, MA  02138

gtroxel@bbn.com

## ABSTRACT
BBN, Harvard, and Boston University are building the DARPA Quantum Network, the world's first network that delivers end-to-end network security via high-speed Quantum Key Distribution, and testing that Network against sophisticated eavesdropping attacks. The first network link has been up and steadily operational in our laboratory since December 2002. It provides a Virtual Private Network between private enclaves, with user traffic protected by a weak-coherent implementation of quantum cryptography. This prototype is suitable for deployment in metro-size areas via standard telecom (dark) fiber. In this paper, we introduce quantum cryptography, discuss its relation to modern secure networks, and describe its unusual physical layer, its specialized quantum cryptographic protocol suite (quite interesting in its own right), and our extensions to IPsec to integrate it with quantum cryptography.

## Categories and Subject Descriptors
C.2.1 [**Network Architecture and Design**]: *quantum cryptography.*

## General Terms
Algorithms, Measurement, Design, Experimentation, Security.

## Keywords
Quantum cryptography, quantum key distribution, secure networks, cryptographic protocols, key agreement protocols, error correction, privacy amplification, IPsec.

## 1. QUANTUM KEY DISTRIBUTION
Modern networks generally rely on one of two basic cryptographic techniques to ensure the confidentiality and integrity of traffic carried across the network: symmetric (secret) key and asymmetric (public) key. Indeed today's best systems generally employ both, using public key systems for

This work is supported by the Defense Advanced Research Projects Agency (DARPA). The views and conclusions contained in this document are those of the authors and should not be interpreted as representing the official policy of DARPA or the U. S. Government.

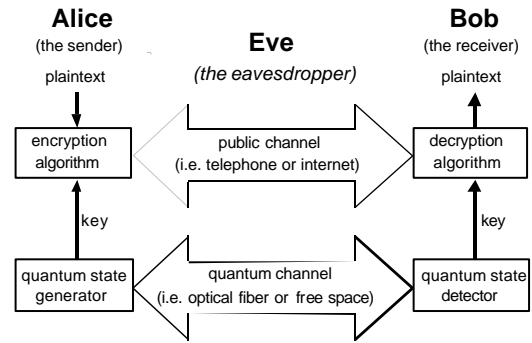

**Figure 1. Quantum Key Distribution.**

authentication and establishment of secret "session" keys, and then protecting all or part of a traffic flow with these session keys. Certain other systems transport secret keys "out of channel," e.g. via courier, as in classical cryptography.

Fundamental aspects of quantum physics – unitarity, the uncertainty principle, and the Einstein-Podolsky-Rosen violation of Bell's inequalities – now suggest a third paradigm for key distribution: quantum cryptography. Initial experiments seem to confirm the utility of this paradigm. Assuming that the theoretical models continue to be confirmed in the use of actual devices, the fundamental laws of nature can be invoked to assure the confidentiality of transmitted data.

As shown in Fig. 1, quantum cryptography – more properly termed Quantum Key Distribution, QKD – employs two distinct channels. One is used for transmission of quantum key material by very dim (single photon) light pulses. The other, public channel carries all message traffic, including the cryptographic protocols, encrypted user traffic, etc.

QKD consists of the transmission of raw key material, e.g., as dim pulses of light from Alice to Bob, via the quantum channel, plus processing of this raw material to derive the actual keys. This processing involves public communication (key agreement protocols) between Alice and Bob, conducted in the public channel, along with specialized QKD algorithms. The resulting keys can then be used for cryptographic purposes, e.g., to protect user traffic. By the laws of quantum physics, any eavesdropper (Eve) that snoops on the quantum channel will cause a measurable disturbance to the flow of single photons. Alice and Bob can detect this, take appropriate steps in response, and hence foil Eve's attempt at eavesdropping.

Quantum cryptography was proposed by Bennett and Brassard in 1984, who also defined the first QKD protocol, called BB84 [1,2]. At time of writing, a handful of research teams around the world have succeeded in building and operating quantum cryptographic devices. Teams at Geneva, Los Alamos, IBM, and elsewhere are performing QKD through telecom fibers [3,4,5]. The best current systems can support distances up to about 70 km through fiber, though at very low bit-rates (e.g. a few bits/second). Teams at Los Alamos and Qinetiq [6,7] are performing free-space quantum cryptography, both through daytime sky and through the night at distances up to 23 km.

In addition to these efforts, whose systems all employ weak-coherent quantum cryptography, there is also interest in cryptography based on a very different physical phenomenon, namely entanglement between pairs of photons produced by Spontaneous Parametric Down-Conversion (SPDC). This form of cryptography has been demonstrated by Geneva [8,9] and discussed in a number of interesting papers [10-16].

The Geneva team has provided a superb overview of current state of the art in QKD [17]. We heartily recommend it to anyone interested in learning more about this fascinating field.

## 2. DESIRABLE QKD ATTRIBUTES

Broadly stated, QKD offers a technique for coming to agreement upon a shared random sequence of bits within two distinct devices, with a very low probability that other devices (eavesdroppers) will be able to make successful inferences as to those bits' values. In specific practice, such sequences are then used as secret keys for encoding and decoding messages between the two devices. Viewed in this light, QKD is quite clearly a key distribution technique, and one can rate QKD's strengths against a number of important goals for key distribution, as summarized in the following paragraphs.

**Confidentiality of Keys.** Confidentiality is the main reason for interest in QKD. Public key systems suffer from an ongoing uncertainty that decryption is mathematically intractable. Thus key agreement primitives widely used in today's Internet security architecture, e.g., Diffie-Hellman, may perhaps be broken at some point in the future. This would not only hinder future ability to communicate but could reveal past traffic. Classic secret key systems have suffered from different problems, namely, insider threats and the logistical burden of distributing keying material. Assuming that QKD techniques are properly embedded into an overall secure system, they can provide automatic distribution of keys that may offer security superior to that of its competitors.

**Authentication.** QKD does not in itself provide authentication. Current strategies for authentication in QKD systems include prepositioning of secret keys at pairs of devices, to be used in hash-based authentication schemes, or hybrid QKD-public key techniques. Neither approach is entirely appealing. Prepositioned secret keys require some means of distributing these keys before QKD itself begins, e.g., by human courier, which may be costly and logistically challenging. Furthermore, this approach appears open to denial of service attacks in which an adversary forces a QKD system to exhaust its stockpile of key material, at which point it can no longer perform authentication. On the other hand, hybrid QKD-public key schemes inherit the possible vulnerabilities of public key systems to cracking via quantum computers or unexpected advances in mathematics.

**Sufficiently Rapid Key Delivery.** Key distribution systems must deliver keys fast enough so that encryption devices do not exhaust their supply of key bits. This is a race between the rate at which keying material is put into place and the rate at which it is consumed for encryption or decryption activities. Today's QKD systems achieve on the order of 1,000 bits/second throughput for keying material, in realistic settings, and often run at much lower rates. This is unacceptably low if one uses these keys in certain ways, e.g., as one-time pads for high-speed traffic flows. However it may well be acceptable if the keying material is used as input for less secure (but often secure enough) algorithms such as the Advanced Encryption Standard. Nonetheless, it is both desirable and possible to greatly improve upon the rates provided by today's QKD technology.

**Robustness.** This has not traditionally been taken into account by the QKD community. However, since keying material is essential for secure communications, it is extremely important that the flow of keying material not be disrupted, whether by accident or by the deliberate acts of an adversary (i.e. by denial of service). Here QKD has provided a highly fragile service to date since QKD techniques have implicitly been employed along a single point-to-point link. If that link were disrupted, whether by active eavesdropping or indeed by fiber cut, all flow of keying material would cease. In our view a meshed QKD network is inherently far more robust than any single point-to-point link since it offers multiple paths for key distribution.

**Distance- and Location-Independence.** In the ideal world, any entity can agree upon keying material with any other (authorized) entity in the world. Rather remarkably, the Internet's security architecture does offer this feature – any computer on the Internet can form a security association with any other, agreeing upon keys through the Internet IPsec protocols. This feature is notably lacking in QKD, which requires the two entities to have a direct and unencumbered path for photons between them, and which can only operate for a few tens of kilometers through fiber.

**Resistance to Traffic Analysis.** Adversaries may be able to perform useful traffic analysis on a key distribution system, e.g., a heavy flow of keying material between two points might reveal that a large volume of confidential information flows, or will flow, between them. It may thus be desirable to impede such analysis. Here QKD in general has had a rather weak approach since most setups have assumed dedicated, point-to-point QKD links between communicating entities which thus clearly lays out the underlying key distribution relationships.

## 3. THE DARPA QUANTUM NETWORK

The DARPA Quantum Network aims to strengthen QKD's performance in these weaker areas. In some instances, this involves the introduction of newer QKD technologies; for example, we hope to achieve rapid delivery of keys by introducing a new, high-speed source of entangled photons. In other instances, we rely on an improved system architecture to achieve these goals; thus, we tackle distance- and location-independence by introducing a network of trusted relays.

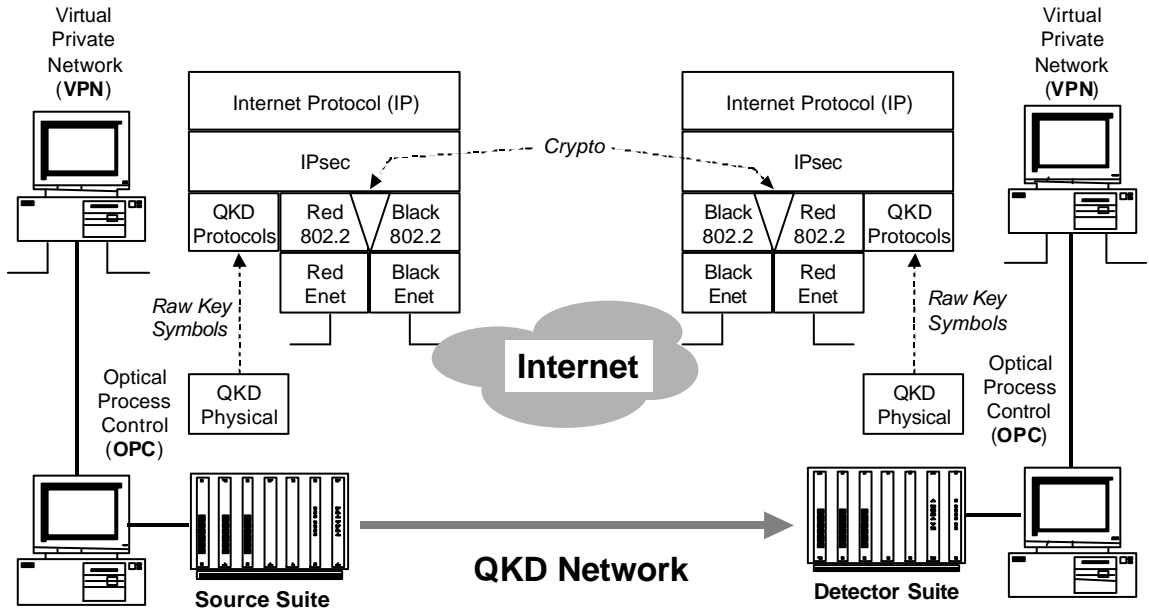

**Figure 2. A Virtual Private Network (VPN) based on Quantum Key Distribution.**

Whereas most work to date has focused on the physical layer of quantum cryptography – e.g. the modulation, transmission, and detection of single photons – our own research effort aims to build QKD *networks*. As such, it is oriented to a large extent towards novel protocols and architectures for highly-secure communications across a heterogenous variety of underlying kinds of QKD links. See [18] for rationale and details of our research plan.

Our security model is the cryptographic Virtual Private Network (VPN). Conventional VPNs use both public-key and symmetric cryptography to achieve confidentiality and authentication/integrity. Public-key mechanisms support key exchange or agreement, and authenticate the endpoints. Symmetric mechanisms (e.g. 3DES, SHA1) provide traffic confidentiality and integrity. Thus VPN systems can provide confidentiality and authentication / integrity without trusting the public network interconnecting the VPN sites.

In our work, existing VPN key agreement primitives are augmented or completely replaced by keys provided by quantum cryptography. The remainder of the VPN construct is left unchanged; see Fig. 2. Thus our QKD-secured network is fully compatible with conventional Internet hosts, routers, firewalls, and so forth.

At time of writing, we are slightly over one year into a projected five-year effort to build the full DARPA Quantum Network. In our first year, we have built a complete quantum cryptographic link, and a QKD protocol engine and working suite of QKD protocols, and have integrated this cryptographic substrate into an IPsec-based Virtual Private Network. This entire system has been continuously operational since December 2002, and we are now in the process of characterizing its behavior and tuning it.

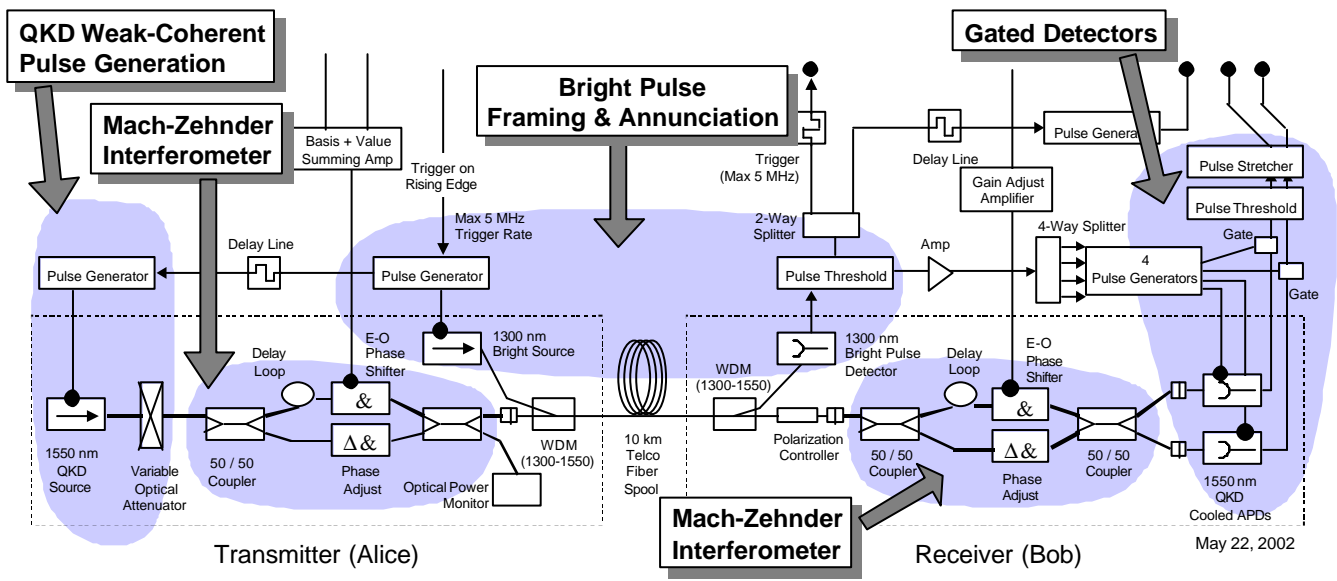

**Figure 3. Our first quantum cryptographic link, based on interferometric phase modulation of single photons.**

In coming years, we plan to build a second link based on two-photon entanglement, and to build various forms of end-to-end networks for QKD across a variety of kinds of links. We expect the majority of our links to be implemented in dark fiber but some may also be implemented in free space, either in the lab or outdoors. Section 8 describes our plans in brief form.

## 4. THE PHYSICAL LAYER

Fig. 3 highlights the major features of our weak-coherent link. As shown, the transmitter at Alice sends single photons by means of a very highly attenuated laser pulse at 1550 nm. Each of these photons passes through a Mach-Zehnder interferometer at Alice which is randomly modulated to one of four phases, thus encoding both a basis and a value in that photon's self interference. The receiver at Bob contains another Mach-Zehnder interferometer, randomly modulated to one of two phases to select a basis. The received photons pass through Bob's interferometer to strike one of the two thermo-electrically cooled single-photon detectors and hence to present a received value. Alice also transmits bright pulses at 1300 nm, multiplexed over the same fiber, to send timing and framing information to Bob.

Figs. 4-6 illustrate the basic mechanism underlying our phase-encoding scheme for conveying qubits. As shown, Alice contains an unbalanced Mach-Zehnder interferometer, i.e., an interferometer in which the two arms have different delays. Bob contains a similar interferometer; in fact, certain dimensions of the two interferometers must be kept identical to within a fraction of the QKD photon's wavelength, i.e., a fraction of 1550 nm. We have labeled the various paths that a photon can follow through these interferometers for ease of discussion in the following paragraphs.

Fig. 5 shows how a single photon behaves as its pulse proceeds from the 1550 nm QKD source at Alice towards the pair of detectors at Bob. Here one should visualize the photon as a wave rather than as a particle. Thus it follows both paths of each interferometer rather than having to choose a single path. Not surprisingly, the part of the photon pulse that follows the shorter arm of an interferometer emerges sooner than that part of the pulse that takes the longer arm. Reading from the left of Fig. 5, we see a single photon pulse emitted from the QKD source. It follows both arms in Alice's interferometer and the part that follows the longer path (labeled $L_A$) begins to lag behind that which takes the shorter path ($S_A$). These two halves are combined at at the 50 / 50 coupler before they leave Alice and travel as two distinct pulses through the telco fiber loop.

Once this double-pulse photon reaches the interferometer at Bob, it once again takes both paths through Bob. Thus the part of the double pulse that takes the top path (long path) will be delayed relative to that which follows the shorter, bottom path.

Fig. 6 shows how Bob's 50 / 50 coupler (just before the detectors) combines the resulting double pulses. If the interferometers are set correctly, the leading pulse in the upper train will align more or less precisely with the trailing pulse in the bottom train, and the two amplitudes will be summed. The right part of the diagram shows the resulting combined waveform at the 50 / 50 coupler just in front of Bob's pair of QKD detectors.

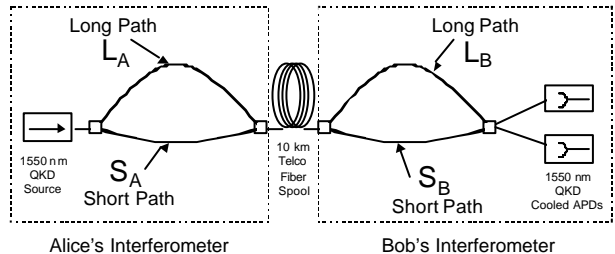

**Figure 4. Paths through unbalanced Mach-Zehnder interferometers.**

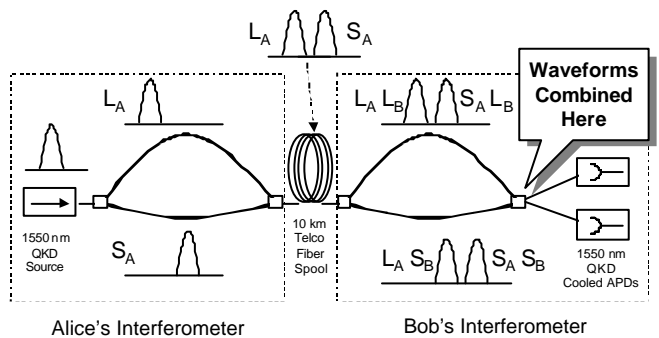

**Figure 5. Effects of an unbalanced interferometer on a photon.**

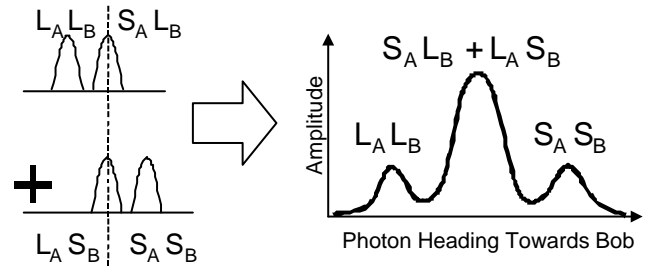

**Figure 6. Recombined photon at 50 / 50 coupler just before Bob's QKD detectors. The central peak is self-interfering.**

We are now, finally, in a position to explain exactly how '0' and '1' values are sent via QKD pulses between Alice and Bob. As we mentioned, this central peak emerged from the combination of double pulse assuming that Alice's and Bob's interferometers were aligned more or less precisely. This is in fact where the (basis, value) modulation enters the picture. First, we need a few basic facts from optics:

- When a light ray is incident on a surface and the material on the other side of the surface has a *higher* index of refraction (i.e. a lower speed of light), then the reflected light ray is phase-shifted by exactly half a wavelength.

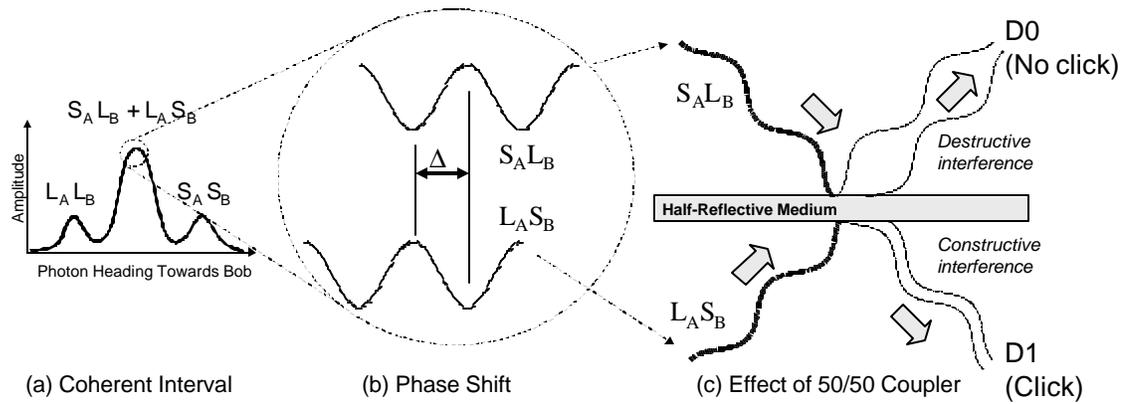

**Figure 7. Signaling '0' and '1' symbols via partial-wavelength phase-shifting of Mach-Zehnder interferometers.**

- When a ray is incident on a surface and the material on the other side has a *lower* index of refraction, the reflected light ray does not have its phase changed.

- When a ray goes from one medium into another, its direction changes due to refraction but no phase change occurs at the surfaces of the two mediums.

- When a ray travels through a medium, such as a glass plate, its phase is shifted by an amount that depends on the index of refraction of the medium and the path length of the ray through the medium.

With these facts as background, we can see how '1' and '0' symbols work in a phase-shifted QKD system. Fig. 7 provides a schematic overview of how the system works. As depicted, the central peak within a photon pulse contains a coherent interval (a) during which two distinct wave paths are present simultaneously. A close-up view of these waves, as in (b), shows that in general the two distinct waves have different phases – that is, the phase of the wave as it traveled through the $S_A L_B$ path is offset by some phase shift, $\Delta$, from that which traveled through the $L_A S_B$ path. At the right, these two waves interact with the final 50/50 coupler in the system to present constructive interference for one detector (click) but destructive interference for the other (no click).

Thus Alice can signal '0' and '1' symbols to Bob merely by adjusting the relative phases of these two waves, i.e, by adjusting the phase delta value ($\Delta$) on a per-pulse basis. Alice does this by setting her phase shifter accordingly for each transmitted pulse.

The phase-encoded variant of BB84 works as follows. Alice encodes the 0 or 1 value for a single photon in either of two randomly selected non-orthogonal bases. She represents the 0 value by either the phase shift of 0 (basis 0) or $\pi/2$ (basis 1), and represents the 1 value by either $\pi$ (basis 0) and $3\pi/2$ (basis 1). Thus Alice can apply one of four phase shifts (0, $\pi/2$, $\pi$, $3\pi/2$) to encode four different corresponding (basis, value) pairs of the key as (00, 01, 10, 11). This is achieved by applying four different voltages to the Electro-Optical (E-O) phase shifter on the transmitter side. It can be seen that the voltage on the phase shifter can be derived as a sum of basis and value bits via a summing amplifier.

When their phase difference is equal to 0 or $\pi$, Alice and Bob are using compatible bases and obtain nominally identical results (assuming a number of unrealistic assumptions such as zero noise, no photon loss, etc.). In such cases, Alice can infer from the phase shift she applied the detector hit at the Bob's end, and hence the bit value Bob registered. By the same process of logic, Bob can deduce which value Alice transmitted. However, when Alice and Bob didn't randomly agree on the same basis (i.e. when their phase difference equals $\pi/2$ or $3\pi/2$), the bases are incompatible and the photon strikes one of the two APDs at random.

Alice provides the clock source (trigger) for both transmitter and receiver. All clocking in this system ultimately derives from a single trigger supplied from higher layers in the protocol stack, and drawn in Fig. 3 as descending from above the Transmitter suite. This clock comes from a FIFO card on a real-time Optical Process Control (OPC) computer not discussed in this paper.

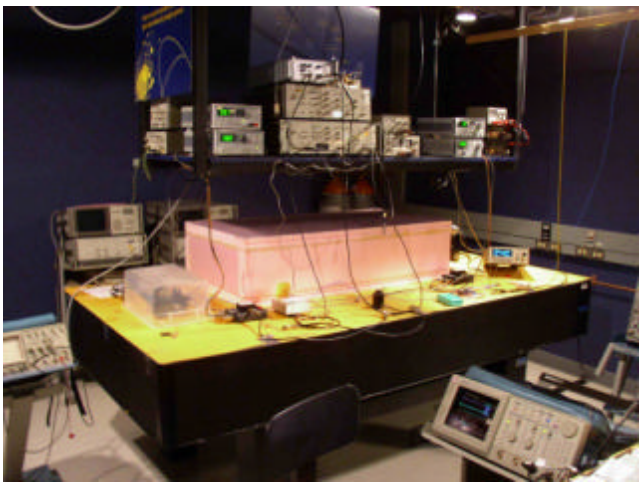
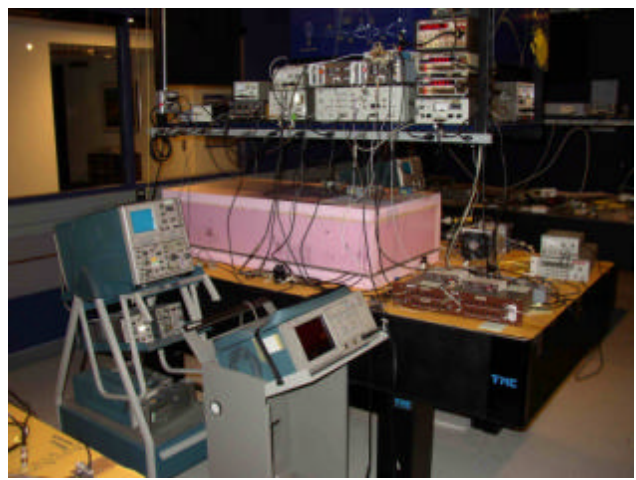

**Figure 8. Physical layers of our weak-coherent link in the laboratory (Alice at left, Bob at right).**

The rising edge of this signal serves as a trigger for the transmitter suite. It triggers a pulse generator that in turn emits two further triggers: one leads immediately to the 1300 nm bright source laser, and the other is delayed and then given to the 1550 nm QKD source laser. Thus the 1300 nm bright pulse laser transmits first, followed a short time later by a dim (single photon) pulse from the 1550 nm QKD source laser. This fixed delay is designed to allow sufficient time for Bob to gate his QKD detectors after receiving the bright pulse.

On the receiver side, Bob's pair of 1550 nm QKD detectors are operated in the Geiger gated mode, where the applied bias voltage exceeds the breakdown voltage for a very short period of time when a photon is expected to arrive, leading an absorbed photon to trigger an electron avalanche consisting of thousands of carriers. Since the typical gating interval is a few ns, this mode of operation requires some knowledge of the photon arrival time, which is deduced from the 1300 nm bright pulse (synchronization) laser and passively quenched sync detector, which generate and detect the trigger signal for the gated APD detectors after a known delay.

At Bob, the received annunciator pulse from the 1300 nm Bright Pulse Detector triggers the gating of Bob's cooled APDs to set the detectors' bias voltages high just around the time that the 1550 nm QKD photon arrives. Bob interprets a click on APD Detector 0 (D0) as a bit value of "0", and on Detector 1 (D1) as "1". After Bob samples these detectors, it then sets up for the next incoming QKD photon by randomly applying a phase shift of either 0 or $\pi/2$ to the Phase Shifter.

Actively controlled fiber stretchers are required in order to stabilize path length during transmission and to maintain the equivalence of interferometers on both sides, with the same coupling ratios in each arm and the same path length. Moreover, the optical components, such as phase shifters and phase adjusters, are polarization dependent, requiring a polarization-maintaining fiber for both interferometers and active polarization controller on the receiver side to restore polarization after passing regular telecom fiber.

At time of writing, our weak-coherent link is operating with a 1 MHz pulse repetition rate, mean photon-emission number of 0.1 photons per pulse, and approximately a 6-8% Quantum Bit Error Rate (QBER) on the detectors cooled to –30 C.

## 5. QKD PROTOCOLS IMPLEMENTATION

Quantum cryptography involves a surprisingly elaborate suite of specialized protocols, which we term "QKD protocols." Many aspects of these protocols are unusual – both in motivation and in implementation – and may be of interest to specialists in communications protocols.

This section describes the protocols now running in our C language QKD protocol implementation. We have designed this engine so it is easy to "plug in" new protocols, and expect to devote considerable time in coming years to inventing new QKD protocols and trying them in practice. As shown in Fig. 5, these protocols are best described as sub-layers within the QKD protocol suite. Note, however, that these layers do not correspond in any obvious way to the layers in a communications stack, e.g., the OSI layers. As will be seen, they are in fact closer to being pipeline stages.

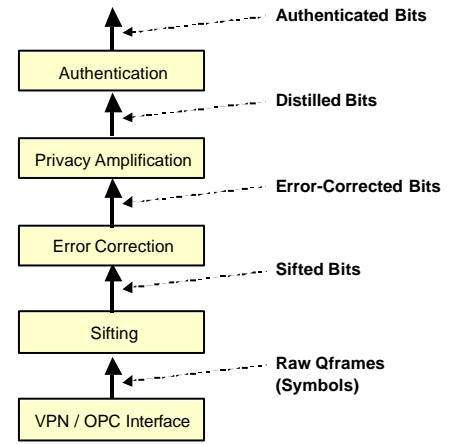

**Fig. 9. The QKD protocol stack.**

**Sifting** is the process whereby Alice and Bob winnow away all the obvious "failed qubits" from a series of pulses. As described in the introduction to this section, these failures include those qubits where Alice's laser never transmitted, Bob's detectors didn't work, photons were lost in transmission, and so forth. They also include those symbols where Alice chose one basis for transmission but Bob chose the other for receiving.

At the end of this round of protocol interaction – i.e. after a sift and sift response transaction – Alice and Bob discard all the useless symbols from their internal storage, leaving only those symbols that Bob received and for which Bob's basis matches Alice's.

In general, sifting dramatically prunes the number of symbols held in Alice and Bob. For instance, assume that 1% of the photons that Alice tries to transmit are actually received at Bob and that the system noise rate is 0. On average, Alice and Bob will happen to agree on a basis 50% of the time in BB84. Thus only 50% x 1% of Alice's photons give rise to a sifted bit, i.e., 1 photon in 200. A transmitted stream of 1,000 bits therefore would boil down to about 5 sifted bits.

**Error correction** allows Alice and Bob to determine all the "error bits" among their shared, sifted bits, and correct them so that Alice and Bob share the same sequence of error-corrected bits. Error bits are ones that Alice transmitted as a 0 but Bob received as a 1, or vice versa. These bit errors can be caused by noise or by eavesdropping.

Error correction in quantum cryptography has a very unusual constraint, namely, evidence revealed in error detection and correction (e.g. parity bits) must be assumed to be known to Eve, and thus to reduce the hidden entropy available for key material. As a result, there is very strong motivation to design error detection and correction codes that reveal as little as possible in their public control traffic between Alice and Bob.

Our first approach for error correction is a novel variant of the Cascade protocol [19] and algorithms. The protocol is adaptive, in that it will not disclose too many bits if the number of errors is low, but it will accurately detect and correct a large number of errors (up to some limit) even if that number is well above the historical average.

Our version works by defining a number of subsets (currently 64) of the sifted bits and forming the parities of each subset. In the first message, the list of subsets and their parities is sent to the other side, which then replies with its version of the parities. The subsets are pseudo-random bit strings, from a Linear-Feedback Shift Register (LFSR) and are identified by a 32-bit seed for the LFSR. Once an error bit has been found and fixed, both sides inspect their records of subsets and subranges, and flip the recorded parity of those that contained that bit. This will clear up some discrepancies but may introduce other new ones, and so the process continues.

Since these parity fields are revealed in the interchange of "error correction" messages between Alice and Bob, these bits must be taken as known to Eve. Therefore, the QKD protocol engine records amount of information revealed (lost) due to parity fields, and later requires a compensating level of privacy amplification to reduce Eve's knowledge to acceptable levels.

**Privacy amplification** is the process whereby Alice and Bob reduce Eve's knowledge of their shared bits to an acceptable level. This technique is also often called advantage distillation.

The side that initiates privacy amplification chooses a linear hash function over the Galois Field $GF[2^n]$ where $n$ is the number of bits as input, rounded up to a multiple of 32. He then transmits four things to the other end—the number of bits $m$ of the shortened result, the (sparse) primitive polynomial of the Galois field, a multiplier ($n$ bits long), and an $m$-bit polynomial to add (i.e. a bit string to exclusive-or) with the product. Each side then performs the corresponding hash and truncates the result to $m$ bits to perform privacy amplification.

**Authentication** allows Alice and Bob to guard against "man in the middle attacks," i.e., allows Alice to ensure that she is communicating with Bob (and not Eve) and vice versa. Authentication must be performed on an ongoing basis for all key management traffic, since Eve may insert herself into the conversation between Alice and Bob at any stage in their communication.

The original BB84 paper [1] described the authentication problem and sketched a solution to it based on universal families of hash functions, introduced by Wegman and Carter [20]. This approach requires Alice and Bob to already share a small secret key, which is used to select a hash function from the family to generate an authentication hash of the public correspondence between them. By the nature of universal hashing, any party who didn't know the secret key would have an extremely low probability of being able to forge the correspondence, even an adversary with unlimited computational power. The drawback is that the secret key bits cannot be re-used even once on different data without compromising the security. Fortunately, a complete authenticated conversation can validate a large number of new, shared secret bits from QKD, and a small number of these may be used to replenish the pool.

There are many further details in a practical system which we will only mention in passing, including symmetrically authenticating both parties, limiting the opportunities for Eve to force exhaustion of the shared secret key bits, and adapting the system to network asynchrony and retransmissions. Another important point: it is insufficient to authenticate just the QKD protocols; we must also apply the these techniques to authenticate the VPN data traffic.

## 6. A DISQUISITION ON EVE

Within the quantum cryptographic community, Eve is generally understood to be limited only by the known laws of physics, and to otherwise possess engineering and mathematical powers far beyond the current state of the art. In particular, it is axiomatic that Eve can:

- Instantly break all symmetric ciphers and public key primitives.
- Detect all dim pulses with zero loss.
- Create dim pulses that are indistinguishable from Alice's except for the limitations of quantum physics (e.g. the no-cloning law).
- Transport photons to Bob with zero loss.
- Eavesdrop undetectably on the public channel.
- Forge or block messages on the public channel.

It will be seen that, given these basic axioms, Eve can launch highly formidable eavesdropping and "man in the middle" attacks against Alice and Bob since she can interpose herself along the photonic channel between Alice and Bob in ways that are very hard to detect, and instantly defeat all ciphers on the "public channel" except one-time pads. Defense against such extreme attacks plays a major role in the theoretical security of a full quantum cryptographic system.

Privacy amplification depends on having an estimate of the eavesdropping-free entropy of the quantum channel—the amount of information in the channel beyond what Eve might know. The estimate is made after sifting and error correction, and any randomness or bias testing. The inputs to entropy estimation are:

$b$, the number of received bits (sifted)
$e$, the number of errors in the sifted bits
$n$, the total number of bits transmitted
$d$, the number of parity bits disclosed during error correction
$r$, a non-randomness measure from randomness tests

The components of the entropy estimate are:

- An estimate of the information Eve possesses due to non-transparent (error-inducing) observations.
- An estimate of the information Eve might possess due to transparent eavesdropping—observations that have no effect on the error rate, e.g. beamsplitting attacks, interceptions of multi-photon pulses, and the like.
- The amount of information disclosed publicly during error detection and correction.
- An estimate of the information Eve might possess due to non-randomness in the raw QKD bits (detector bias, for example).

Of these components, only the third—publicly disclosed information—is clear and non-controversial: it is precisely the number of sets of bits whose parities have been disclosed. The fourth—the non-randomness measure—is only a placeholder at the moment, until randomness testing is put into the system. We assume that this testing will produce an measure in the form of a number of bits by which to shorten the string.

Information from transparent eavesdropping is not uniformly treated in the QKD community. This category includes all eavesdropping that doesn't cause errors, which was originally thought to include only beamsplitting attacks on multi-photon pulses. It is now becoming clear that there are more general attacks of the same ilk. For instance, Brassard et al. [13] have pointed out that all weak coherent systems are particularly vulnerable to so-called "Positive Operator Valued Measure" attacks. The amount of information leakage can be proportional to the number of transmitted bits times the multi-photon probability, rather than the number of bits received by Bob. With an entangled-photon link, by contrast, the amount of information Eve may obtain is only proportional to the number of received bits times the multi-photon probability.

An unresolved issue is the amount of information Eve obtains due to non-transparent eavesdropping. Following Slutsky et al. [21], we call these defense functions. Several defense functions have been published for quantum cryptographic systems. Two of the best known are by Bennett, et al. [1] and Slutsky et al [21]; see Appendix for details. Neither appears to be completely accurate—Bennett's estimate does not take into account all the information Eve can get from indirect attacks that give an error rate less than 25%, and while Slutsky's estimate may be asymptotically correct, it is overly conservative for finite-length blocks.

Because we expect to refine entropy estimates in future, and perform comparisons of different systems under like assumptions, we provide a choice of defense function. At present, we support both Bennett and Slutsky. Both estimates include a margin for certainty based on the standard deviation. In Bennett's estimate, this is 5 standard deviations, including the standard deviation of the multi-photon probability. In Slutsky's case it is parameterizable in terms of probability of a successful attack, but doesn't include multi-photon probabilities. For consistency, we separate out the standard deviation of each term and combine them at the end, times a confidence parameter $c$ (so a parameter $c = 5$ mean 5 standard deviations, or about $10^{-6}$ chance of successful eavesdropping).

## 7. IPSEC EXTENSIONS

Once "finished" quantum cryptographic key material is available, it can be employed as keys for one or more applications. Our first use of such key material is as standard cryptographic keys for IPsec-based Virtual Private Networks. We have accordingly extended both the IPsec traffic processing path and its key agreement protocol (IKE) to employ key material obtained by quantum cryptography.

Because new key material is constantly streaming into both Alice and Bob, they can both update the keys used in their cryptographic algorithms more or less continually. At present we use these keys as input to the IPsec Phase 2 hash, and update the resultant AES keys about once a minute.

IPsec is an architectural framework for secure communications within the Internet Protocol suite. This framework is defined by a standards-track document within the Internet Engineering Task Force (IETF), namely RFC 2401, Security Architecture for the Internet Protocol. One of its components, the Internet Key Exchange (IKE) protocol, permits two endpoints to agree first on which cryptographic protocols and algorithms they wish to employ for a given security association, and second on the keys they use to encrypt and/or authenticate subsequent message traffic within that security association. IKE is defined by its own standards-track document within the IETF, RFC 2409, The Internet Key Exchange (IKE).

Although IKE is a relatively complicated protocol, its basic concepts are straightforward. Fig. 11 depicts the most important elements involved in an ongoing relationship between two IKE peers. This illustration is intended to be high-level and schematic rather than a detailed depiction of an actual software architecture.

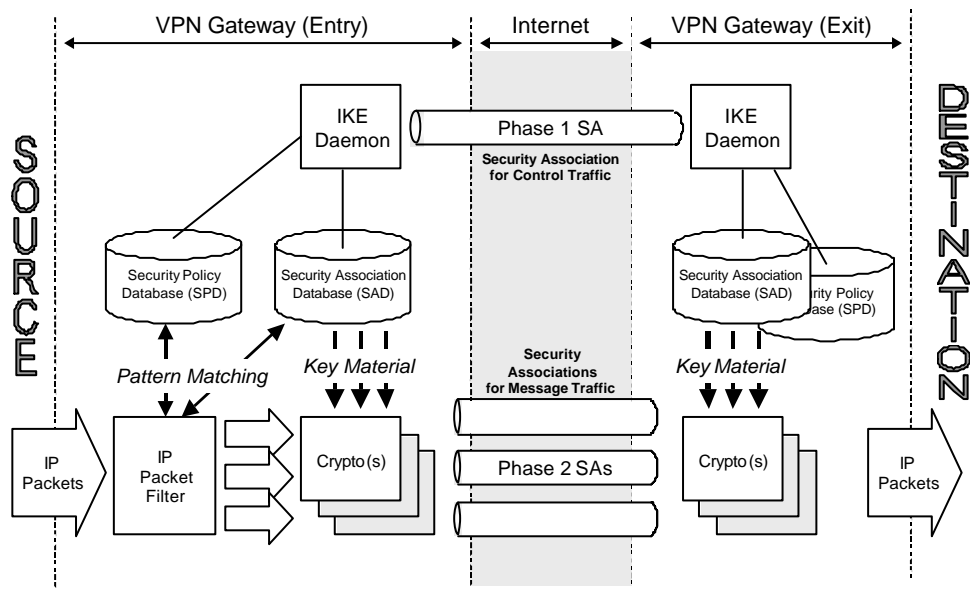

**Figure 10. Simplified schematic of the IKE / IPsec architecture for Virtual Private Networks.**

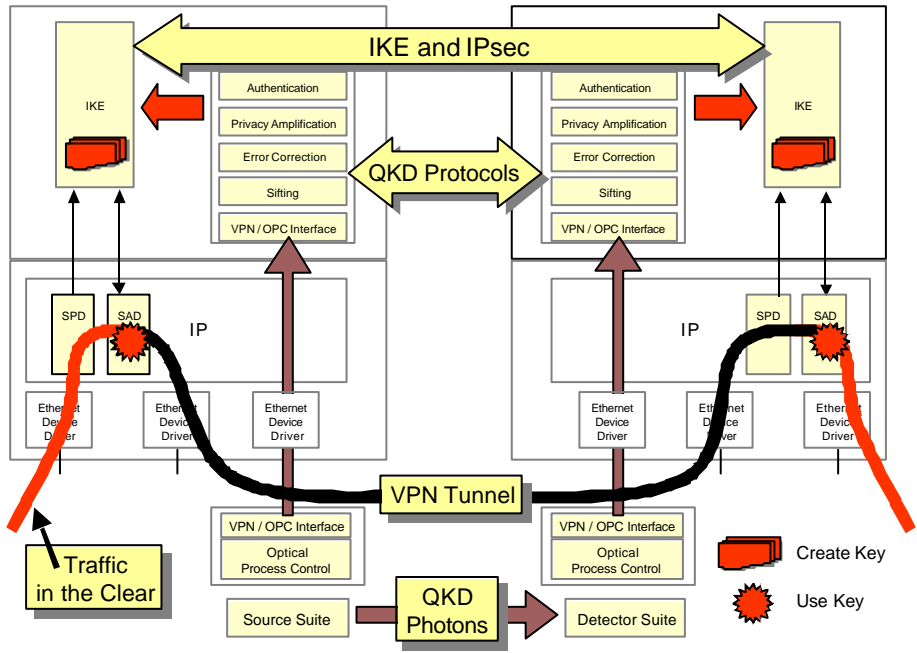

**Figure 11. Our architecture for Virtual Private Networks based on quantum cryptography.**

Every security association has a maximum lifetime which governs how long the key material for that association can be used. This lifetime can be expressed either in time (seconds) or in data encrypted (kilobytes) and is configured via the Security Policy Database (SPD) entry for a given security association. Every time the lifetime expires, a new security association must be negotiated and it will bring with it fresh key material. This is sometimes termed "key rollover," because it replaces an older key by a newer one while still protecting the same underlying traffic flow.

The key size needed for an encryption or hash algorithm depends on the specific algorithm being employed. Some algorithms always use the same key size. For others, which can accept a range of key sizes, the Security Policy Database (SPD) must list the actual key size that should be employed for a given SA.

Our first implementation runs in NetBSD with the 'raccoon' IKE daemon. We have modified both the kernel and the IKE daemon to introduce QKD extensions. (The QKD protocol engine also runs on the same NetBSD platform.)

Our extensions to date fall into two basic categories: those that use QKD techniques for agreement on secret keys that are then employed as seeds for conventional symmetric ciphers (e.g. AES and 3DES) with continual and automatic reseeding by fresh QKD bits, and those that use a sequence of QKD bits as a one-time pad or Vernam cipher for the message traffic.

In the **rapid-reseeding** extension, we have included distilled QKD bits into the IKE Phase 2 hash, so that keys protecting IPsec Security Associations (SAs) are derived from QKD. Fig. 12 shows the very first time this extension worked with real QKD bits, i.e., the first time to our knowledge that a VPN was created that protected its traffic with quantum cryptography.

In the **one-time pad** extension, we have introduced one-time pad mechanisms into IKE and IPsec traffic processing, with all the adjutant changes to protocols and algorithms required, and the necessary linkages into the secret bits obtained via QKD.

Ancillary changes include policy mechanisms to specify when either of these extensions should be used, on a per-tunnel basis, and negotiation mechanisms to agree on which QKD bits will be used. Thus our implementation currently supports multiple VPNs per cryptographic gateway, each with its own set of cryptographic algorithms, keys, rekey rates, and so forth. Some may use conventional cryptography (e.g. AES), while others employ one-time pads, depending on how sensitive traffic is within a given VPN.

Use of IKE for quantum cryptography calls special attention to a few rarely-exercised parts of the IKE design that may have some impact on our overall system design.

The first such aspect concerns timeouts, in particular, the maximal amount of time that can elapse during an IKE negotiation. Such values are often set to 10s of seconds for Phase 1 negotiation, and less than 10 seconds for Phase 2. These values may be too small for systems employing QKD since it may take a while to accumulate enough bits for a successful negotiation. In addition, of course, this narrow window makes Eve's denial-of-service attacks somewhat easier since she must block IKE messages during only a relatively short time in order to bring down the security association(s).

The second concerns what IKE does when Alice and Bob believe they possess secret bits in common but in fact these two sets of bits are not identical. This may well happen in quantum cryptography, since noise on the single-photon channel can only be detected and corrected probabilistically. As it happens, IKE has no mechanisms for noticing or dealing with such cases. The result appears to be that all security associations that employ key bits derived from this corrupted information will fail to properly encrypt / decrypt traffic. This situation will apparently continue until the security association is renewed, i.e., rolls over to a new security association. This aspect of IKE provides some pressure for adjusting the QKD error correction protocols towards a low residual bit error rate.

```
Dec  5 12:53:32 bob-gw racoon: INFO: isakmp.c:1046:isakmp_ph2begin_r(): respond new phase 2 negotiation: 192.1.99.35[0]<=>192.1.99.34[0]
Dec  5 12:53:32 bob-gw racoon: INFO: proposal.c:1023:set_proposal_from_policy(): RESPONDER setting QPFS encmodesv 1
Dec  5 12:53:32 bob-gw racoon: INFO: bbn-qkd-qpd.c:1047:qke_create_reply(): reply 1 Qblocks 1024 bits 1024.000000 entropy (offer is 1 Qblocks)
Dec  5 12:53:32 bob-gw racoon: INFO: oakley.c:473:oakley_compute_keymat_x(): KEYMAT using 128 bytes QBITS
Dec  5 12:53:32 bob-gw racoon: INFO: oakley.c:473:oakley_compute_keymat_x(): KEYMAT using 128 bytes QBITS
Dec  5 12:53:32 bob-gw racoon: INFO: pfkey.c:1107:pk_recvupdate(): IPsec-SA established: ESP/Tunnel 192.1.99.34->192.1.99.35 spi=163084584(0x9b87928)
Dec  5 12:53:32 bob-gw racoon: INFO: pfkey.c:1319:pk_recvadd(): IPsec-SA established: ESP/Tunnel 192.1.99.35->192.1.99.34 spi=136597565(0x824503d)
Dec  5 12:53:43 bob-gw racoon: INFO: pfkey.c:1365:pk_recvexpire(): IPsec-SA expired: AH/Transport 192.1.99.35->192.1.99.36 spi=30999473(0x1d903b1)
Dec  5 12:53:43 bob-gw racoon: INFO: isakmp.c:939:isakmp_ph2begin_i(): initiate new phase 2 negotiation: 192.1.99.35[0]<=>192.1.99.36[0]
```

**Figure 12. Extract from the first IKE transaction setting up a VPN protected by quantum cryptography. Traffic flowed a few moments later.**

Finally we note that our QKD work is not closely tied to IKE itself. It is readily portable to IKEv2, JFK, or indeed upper-layer protocols such as SSL in short order.

## 8. PLANS FOR THE NEXT STEPS

We are now starting to build multiple QKD links woven into an overall QKD *network* that connects its QKD endpoints via a mesh of QKD relays or routers. When a given point-to-point QKD link within the relay mesh fails – e.g. by fiber cut or too much eavesdropping or noise – that link is abandoned and another used instead. This emerging DARPA Quantum Network can be engineered to be resilient even in the face of active eavesdropping or other denial-of-service attacks.

Such QKD networks can be built in several ways. In one variant, the QKD relays may only transport keying material. After relays have established pairwise agreed-to keys along an end-to-end point, e.g., between the two QKD endpoints, they may employ these key pairs to securely transport a key "hop by hop" from one endpoint to the other, being onetime-pad encrypted and decrypted with each pairwise key as it proceeds from one relay to the next. In this approach, the end-to-end key will appear in the clear within the relays' memories proper, but will always be encrypted when passing across a link. Such a design may be termed a "key transport network."

Alternatively, QKD relays may transport both keying material and message traffic. In essence, this approach uses QKD as a link encryption mechanism, or stitches together an overall end-to-end traffic path from a series of QKD-protected tunnels.

Such QKD networks bring important benefits that greatly mitigate the drawbacks of point-to-point links enumerated at the start of this section. First, they can extend the geographic reach of a network secured by quantum cryptography, since wide-area networks can be created by a series of point-to-point links bridged by active relays. Links can be heterogeneous transmission media, i.e., some may be through fiber while others are freespace. Thus in theory such a network could provide fully global coverage. Second, they lessen the chance that an adversary could disable the key distribution process, whether by active eavesdropping or simply by cutting a fiber. A QKD network can be engineered with as much redundancy as desired simply by adding more links and relays to the mesh. Third, QKD networks can greatly reduce the cost of large-scale interconnectivity of private enclaves by reducing the required (N x N-1) / 2 point-to-point links to as few as N links in the case of a simple star topology for the key distribution network.

Such QKD networks do have drawbacks, however. Their prime weakness is that the relays must be *trusted*. Since keying material and – directly or indirectly – message traffic are available in the clear in the relays' memories, these relays must not fall into an adversary's hands. They need to be in physically secured locations and perhaps guarded if the traffic is truly important. In addition, all users in the system must trust the network (and the network's operators) with all keys to their message traffic. Thus a pair of users with unusually sensitive traffic must expand the circle of those who can be privy to it to include all machines, and probably all operators, of the QKD network used to transport keys for this sensitive traffic.

As in classical cryptography, an end-to-end approach is likely to provide the most satisfactory architecture for disentangling the users' keying material for secured traffic flows from the network that transports such flows. Hence we are building unamplified photonic switches into our QKD network architecture in order to provide end-to-end key distribution via a novel mesh of *untrusted* switches.

Untrusted QKD switches do not participate in QKD protocols at all. Instead they set up all-optical paths through the network mesh of fibers, switches, and endpoints. Thus a photon from its source QKD endpoint proceeds, without measurement, from switch to switch across the optical QKD network until it reaches the destination endpoint at which point it is detected. We currently anticipate that the QKD switches will be built from MEMS mirror arrays, or equivalents, together with novel distributed protocols and algorithms that allow end-to-end path setup across the network, and that (as in untrusted networks) provide a robust means for routing around eavesdropping or failed links.

Untrusted QKD networks have different strengths and weaknesses than trusted QKD networks. Their main strength is that support truly end-to-end key distribution; QKD endpoints need not share any secrets with the key distribution network or its operators. This feature may be extremely important for highly secure networks. Their weaknesses appear significant, however. Unlike trusted relays, untrusted switches cannot extend the geographic reach of a QKD network. In fact, they may significantly reduce it since each switch adds at least a fractional dB insertion loss along the photonic path. In addition, it will also prove difficult in practice to employ a variety of transmission media within an untrusted network, since a single frequency may not work well along a composite path that includes both fiber and freespace links. Untrusted networks may also introduce new vulnerabilities to traffic analysis.

Looking to the later years of the DARPA Quantum Network, the principal weakness in untrusted QKD networks – limited geographic reach – may perhaps be countered by quantum repeaters. There is now a great deal of active research aiming towards such repeaters, and if practical devices are ever achieved, they should slide neatly into the overall architecture of untrusted QKD networks to enable seamless QKD operations over much greater distances than currently feasible.

## 9. CONCLUSIONS

The DARPA Quantum Network demonstrates that quantum cryptography may indeed be used, in practice, to provide continuous key distribution for Internet virtual private networks. However certain critical aspects of the theory of quantum cryptography are still very murky. These include the variety of possible attacks and the detailed quantum mechanical theory underlying photon production, propagation, detection, and so forth. In short, it is now clear that quantum cryptography is feasible in *practice* – but the question still remains as to whether it's feasible in *theory*. Thus our network manifestly works, but it may not truly be secure!

Accordingly, we believe that future work should proceed on two tracks. First, detailed security analyses must be carried out – ranging all the way from quantum mechanical analyses to more traditional network security analyses. Second, work should proceed at full speed on building out the next portions of the network – its next kinds of QKD links (based on entangled photon pairs), and its highly specialized switches.

## 10. ACKNOWLEDGEMENTS

We are deeply indebted to Dr. Mike Foster (DARPA) and Dr. Don Nicholson (Air Force Research Laboratory) who are the sponsor and agent, respectively, for this project. This paper reflects highly collaborative work within the project team. Of these, particular credit is due to Alexander Sergienko and Gregg Jaeger (Boston University), John Myers and Tai Wu (Harvard), and Alex Colvin, William Nelson, Oleksiy Pikalo, John Schlafer, and Henry Yeh (BBN). Our interest in QKD networks was sparked by the prior work of, and discussions with, the quantum cryptography groups at IBM Almaden and Los Alamos, and by the kind hospitality of Dr. David Murley several years ago.

APPENDIX

| QKD Sub-Function | Specific Techniques in the DARPA Quantum Network | Description |
|---|---|---|
| Authentication | Universal Hash Function [*Not yet implemented*] | Preposition a "small" shared secret key at Alice and Bob, and use this key as input to a Universal Hash function along with the refined bits obtained by QKD protocols. Use the result as a cryptographic checksum to verify Alice or Bob's identify. |
| | Hybrid Public Key / Universal Hash Function | Combine the Universal Hash Function approach with Public Key Cryptography, i.e., digital signatures. While this approach does not satisfy the purist quantum cryptographic community, it may be a good engineering approach for communities that believe that public key techniques are still useful. |
| Estimation of Eve's Knowledge | Bennett Estimate | $\frac{4e}{\sqrt{2}}$, std. deviation $\sqrt{(4+2\sqrt{2})e}$. See Bennet et al [1] |
| | Slutsky Estimate | $t = (b-e)\left[1 + \log_2\left(1 - \frac{1}{2}\left(\frac{\max(1-3e',0)}{1-e'}\right)^2\right)\right]$, $s = \sqrt{b}$ where $e' = \frac{e}{b} + \frac{c}{\sqrt{b}}$<br>See Slutsky et al [21] |
| | Resultant Entropy | For both Bennett and Slutsky estimates:<br>$b - r - d - t - m_1 n - m_2 b - c\sqrt{s^2 + m_1 n + m_2 b}$ |
| Privacy Amplification | Universal Hash Function based on Rényi Entropy | Use a hash function to reduce the size of a batch of error-corrected, shared secret bits by an amount sufficient to reduce Eve's possible knowledge of the resultant bits' contents to a sufficiently small amount (e.g. far less than 1 bit's worth). |
| Error Correction | Parity Checks | A conventional parity-checking scheme as widely employed in telecommunications systems. |
| | Cascade | Select random subsets of the sifted bits, compute and exchange parity bits on a subset to detect errors, and then use a divide-and-conquer scheme to correct any detected errors. |
| Sifting | Run-Length Encoding | Encode the sifting messages, as sent between Bob and Alice, efficiently so that runs of identical values (and in particular of "no detection" values) are compressed to take very little space. |